\documentclass[12pt]{iopart}

\usepackage{graphicx}
\usepackage{hyperref}

\usepackage{amssymb}
\usepackage{bm}

\def\onlinecite{\cite}

\newcommand{\bea}{\begin{eqnarray}}
\newcommand{\eea}{\end{eqnarray}}
\newcommand{\la}{\label}
\newcommand{\be}{\begin{equation}}
\newcommand{\ee}{\end{equation}}

\begin{document}

\title[Quantum fluctuations \ldots]{Quantum 
fluctuations of one-dimensional free fermions and Fisher-Hartwig formula for Toeplitz
determinants}

\author{Alexander G.~Abanov}
\address{Department of Physics and Astronomy,
Stony Brook University,  Stony Brook, NY 11794-3800.}

\author{Dmitri A.~Ivanov}
\address{Institute of Theoretical Physics,
Ecole Polytechnique F\'ed\'erale de Lausanne (EPFL), 
CH-1015 Lausanne, Switzerland}

\author{Yachao Qian}
\address{Department of Physics and Astronomy,
Stony Brook University,  Stony Brook, NY 11794-3800.}

\date{\today}

\begin{abstract}
We revisit the problem of finding the probability distribution of a
fermionic number of one-dimensional spinless free fermions on a
segment of a given length. The generating function for this probability
distribution can be expressed as a determinant of a Toeplitz matrix.
We use the recently proven generalized Fisher--Hartwig conjecture
on the asymptotic behavior of such determinants to find the generating
function for the full counting statistics of fermions on a line segment. 
Unlike the method of bosonization,
the Fisher--Hartwig formula correctly takes into account the discreteness
of charge. Furthermore, we check numerically the precision of the 
generalized Fisher--Hartwig formula, find that it has a higher precision
than rigorously proven so far, and conjecture the form of the next-order 
correction to the existing formula.
\end{abstract}


\maketitle

\tableofcontents
\title[Quantum fluctuations of one-dimensional free fermions \ldots]{}

\section{Introduction}

The problem of finding the fermion-number fluctuation on a segment for
free one-dimensional spinless fermions is a classical textbook problem.
It is related to a variety of topics in the theory of one-dimensional
quantum systems and in mathematical physics, such as quantum spin chains \cite{LSM-1961,BarouchMcCoy-1971},
bosonization \cite{Stone-bosonization}, 
full counting statistics (FCS) \cite{1993-LevitovLesovik,1997-IvanovLeeLevitov}, random-matrix theory \cite{mehta},
theory of Toeplitz determinants \cite{1998-Aristov,Ehrhardt-2001}, etc. 
Remarkably, some helpful results
related to this simple problem have been obtained only recently: 
a generalized version of the Fisher-Hartwig conjecture on Toeplitz 
determinants has been proven \cite{2010-DeiftItsKrasovsky}. 
These mathematical results were further used in a non-equilibrium
bosonization approach in Refs.~\onlinecite{2010-GutmanGefenMirlin,2011-GutmanGefenMirlin}.

In the present paper, we use the generalized Fisher--Hartwig formula to
calculate the generating function for FCS of free fermions
on a one-dimensional line segment. Unlike the bosonization approach, the generalized
Fisher--Hartwig formula treats correctly the discreteness of particles. As a result,
we obtain a quantitative description of the development of the
singularity in the FCS generating function in the limit of a large segment length
(a FCS phase transition, in the definition of Ref.~\onlinecite{2010-IvanovAbanov}).

Furthermore, we investigate numerically the precision of the existing generalized
Fisher--Hartwig formula and come to the conclusion that it is more precise than
formally proven. In addition, our numerical results allow us to extract the next-order
correction to the existing formula and to conjecture its analytic form.

The paper is organized as follows. In Section~\ref{sec:formulation} we define the
problem of counting free fermions on a one-dimensional line segment. 
In Section~\ref{sec:main}, we briefly review the existing analytical approaches
to the problem and apply the generalized Fisher--Hartwig formula to obtain
a good approximation for the FCS generating function. In Section~\ref{sec:improving},
we analyze the precision of the obtained expression numerically and conjecture
its improvement by including a higher-order correction. Finally, in 
Section~\ref{sec:summary} we summarize our findings. Some technical details
are presented in the Appendices.

\section{Formulation of the problem: counting fermions on 1D line segment}
\label{sec:formulation}

Consider free spinless fermions in one dimension (either on a line or on a 
one-dimensional lattice) at zero temperature. We study the fluctuations
of the fermion number on a segment of a large length $L$. One can easily
find (using either Wick theorem or bosonization) that
\begin{equation}
	\langle\langle Q^{2}\rangle\rangle \equiv \langle 
	(Q-\langle Q\rangle)^{2}\rangle 
	\sim \frac{1}{\pi^{2}} \ln (L/l_0)
 \label{Q2cont}
\end{equation}
in the limit of a large segment length $L \gg l_0$. Here
$Q=\int_{0}^{L}dx\, c^{\dagger} c$ is the operator of the total
number of fermions (total charge) on the segment ($c^\dagger$ and $c$ are
the fermionic creation and annihilation operators). The averaging is
performed over the ground state of free fermions characterized by the
Fermi wave vector $k_{F}$. The ultraviolet cutoff $l_0$ is given by
$l_0 \sim k_F^{-1}$ on the line and by $l_0 \sim (\sin k_F)^{-1}$ on
the lattice. The average charge on the segment is given exactly by
\begin{equation}
	\langle Q\rangle = L k_F/ \pi\, ,
 \la{Q1}
\end{equation}
(so that the average density of fermions is $k_F/\pi$). 

A more complicated problem is to find all moments of charge 
$\langle Q^{k}\rangle$ or, equivalently, the full distribution of
probabilities of having a given charge $q$ on a segment. Instead of
calculating all moments separately, it is convenient to introduce the
characteristic function
\begin{equation}
	\chi(\lambda) \equiv \langle e^{i\lambda Q}\rangle = \sum_{q=0}^\infty P_{q}e^{i\lambda q}
 \la{FCSdef}
\end{equation}
also known as the full-counting-statistics (FCS) generating function. The
charge cumulants may be expressed as its logarithmic derivatives:
\begin{equation}
	\langle\langle Q^{k} \rangle\rangle 
	= (-i\partial_{\lambda})^{k}\log\chi(\lambda)\Big|_{\lambda=0}\, .
 \la{cumdef}
\end{equation}

It is obvious from the definition of (\ref{FCSdef}) that (i)
$\chi(\lambda)$ is normalized so that $\chi(0)=1$, (ii) 
$\chi(-\lambda) = \chi^*(\lambda)$, 
(iii) it is periodic $\chi(\lambda+2\pi)=\chi(\lambda)$. The
latter property follows from the fact that the charge operator is
integer-valued (all charges are integer numbers).
Our goal is to find a good approximation for $\chi(\lambda)$
in the limit of large $L$. 

\section{From bosonization to generalized Fisher--Hartwig conjecture:
early approaches and new results}
\label{sec:main}

The simplest approach to calculating $\chi(\lambda)$ is bosonization,
which is equivalent to
assuming that the density fluctuations are Gaussian. This assumption
leads immediately to the following result
\begin{equation}
	\ln \chi(\lambda) \approx i\lambda\langle Q\rangle
	-\frac{\lambda^{2}}{2}\langle\langle Q^2\rangle\rangle
 \la{chiGauss}
\end{equation}
with the average charge and the variance given by (\ref{Q1}) and (\ref{Q2cont}),
respectively (see \ref{app:bosonization} for a brief review of the bosonization calculation).

One can notice two obvious drawbacks of this approximation. First,
the bosonization method does not allow to calculate the numerical 
coefficient for the ultraviolet cut-off $l_0$. 
This problem can be partly solved by a direct calculation of
the second moment in the original fermionic representation using
the Wick theorem. Such a calculation (see \ref{app:cumulants}) reproduces the result (\ref{Q2cont})
with
\begin{equation}
	l_0^{-1}= 2 e^{\gamma_{E}+1}\sin k_{F}
 \label{cutoff}
\end{equation}
in the lattice problem and $l_0^{-1}= 2 e^{\gamma_{E}+1} k_{F}$ in the
continuous case (note that the formulas for the continuous case may
be obtained from the results on the lattice by taking the limit $k_F \to 0$
while keeping the product $k_F L$ fixed). Here
$\gamma_{E}=0.57721\ldots$ is the Euler-Mascheroni constant.

Nevertheless, even after fixing the numerical coefficient in $l_0$,
the bosonization result for $\chi(\lambda)$ is only precise up to
a numerical $\lambda$-dependent coefficient of order one. These
$\lambda$-dependent corrections may be interpreted as cumulants of
order higher than two, which are not captured by bosonization.
If we include those corrections, we arrive at the following expansion:
\begin{equation}
	\ln \chi(\lambda) = i\lambda \frac{k_F}{\pi} L - 
	\frac{\lambda^2}{2\pi^2} \ln\frac{L}{l_0}
	+ F_0(\lambda) + o(1)
 \label{chi-zero-order}
\end{equation}
as $L \to \infty$. 
The function $F_0(\lambda)$ may, in principle, be
reconstructed from the cumulants $\langle\langle Q^{k} \rangle\rangle$
calculated with the help of the Wick theorem. Such calculations
appear to be very tedious. Fortunately, the exact expression for
$F_0(\lambda)$ can be obtained using the determinant representation of 
(\ref{FCSdef}) and the Fisher--Hartwig formula for Toeplitz
determinants (see \ref{app:FH} for details and references):
\begin{equation}
	F_0(\lambda)= 2 \ln \left| G\left(1+\frac{\lambda}{2\pi}\right)
	G\left(1-\frac{\lambda}{2\pi}\right) \right| + 
	\frac{\lambda^2}{2\pi^2} (\gamma_E + 1)\, ,
 \label{F-Barnes}
\end{equation}
where $G(z)$ is the Barnes $G$-function \cite{DLMF}. Remarkably,
$F_0(\lambda)$ does not depend on $k_F$, even in the lattice
version. \footnote{In the continuum limit this expression is identical to 
Eqs.~(4.6) and (4.19) of Ref.~\cite{2004-CheianovZvonarev} obtained from 
the asymptotics of a Fredholm determinant.} 

The second deficiency of the bosonization approximation is that
$\chi(\lambda)$ does not obey the periodicity property. This
is due to the nature of the bosonization approximation which
treats the fermionic density as a continuous field and thus 
ignores the discreteness of the fermionic charge. This problem
was addressed in Ref.~\cite{1998-Aristov}, where a
phenomenological formula was proposed to restore the
periodicity (taking into account umklapp processes, in the
bosonization language). Aristov's ansatz in Ref.~\cite{1998-Aristov}
corresponds to neglecting the correction
$F_0(\lambda)$ with its nontrivial $\lambda$ dependence.

In the present work, we show that a rigorous formula restoring
the periodicity of $\chi(\lambda)$ can be obtained by an 
application of the recently proven {\em generalized}
Fisher--Hartwig conjecture \cite{2010-DeiftItsKrasovsky}. It turns out that the correct recipe for the 
periodic extension of Eq.~(\ref{chi-zero-order}) is simply to 
add two such expressions with shifted values of $\lambda$ in the
vicinity of the ``switching points'' $\lambda=(2k+1)\pi$:
\begin{equation}
	\chi(\lambda) \approx \chi_0(\lambda - 2k\pi) + 
	\chi_0(\lambda - 2[k+1]\pi)\, ,
 \label{chi-sum}
\end{equation}
where $\ln \chi_0(\lambda)$ is given by the right-hand side of
Eq.~(\ref{chi-zero-order}). The two terms are of the same
order of magnitude at the ``switching point'' 
$\lambda=(2k+1)\pi$, but one of them becomes subleading
(in $L$)
away from this point. For the same reason, only two
shifted values of $\lambda$ are of relevance at each
``switching point'': shifts by higher multiples of $2\pi$
produce terms decaying as higher powers of $L$ and therefore
may be neglected. The details of the application of the
Fisher--Hartwig formula to our problem
are presented in \ref{sec:fh}.

\section{Improving the generalized Fisher--Hartwig
formula: numerical analysis}
\label{sec:improving}

In the proof of the generalized Fisher--Hartwig conjecture in
Ref.~\cite{2010-DeiftItsKrasovsky}, its precision is only
estimated as a relative $o(1)$ as $L \to \infty$. 
This implies that, in the
approximation (\ref{chi-sum}), both terms are within the
proven precision only exactly at the switching point (and therefore,
at the switching point, the expansion (\ref{chi-zero-order}) fails).
Away from the switching point, the subleading term is already
beyond the rigorously proven precision (and thus the expansion
(\ref{chi-zero-order}) is the best proven estimate). 

In order to rectify this situation, we perform a numerical
analysis of the generalized Fischer--Hartwig formula
(\ref{chi-sum}) based on exact evaluation
of Toeplitz determinants of sizes up to $5000$. We claim
that the formula (\ref{chi-sum}) has a higher precision
than that rigorously proven: the subleading of the two
terms in Eq.~(\ref{chi-sum}) provides the main correction
to the asymptotic behavior (\ref{chi-zero-order}). Moreover,
the next-order correction may be captured with
the use of the following conjectured formulas (to
simplify notation, we specify to the interval
$\lambda \in [0, 2\pi]$):
\begin{equation}
	\chi(\lambda) = \chi_1(\lambda) + 
	\chi_1(\lambda - 2\pi) +\varepsilon\, ,
 \label{chi-sum-2}
\end{equation}
where
\begin{equation}
	\chi_1(\lambda)=\exp \left[ i\lambda \frac{k_F}{\pi} L - 
	\frac{\lambda^2}{2\pi^2} \ln\frac{L}{l_0}
	+ F_0(\lambda) + F_1(k_F,\lambda) L^{-1} \right]
 \label{F1-definition}
\end{equation}
and the higher-order correction $\varepsilon$ is of the
relative order $L^{-2}$. More precisely, it can be estimated as
\begin{equation}
	\varepsilon=\Big( |\chi_1(\lambda)| + 
	|\chi_1(\lambda - 2\pi)| \Big) \cdot  O(L^{-2})\, .
 \label{epsilon-estimate}
\end{equation}

We confirm this conjecture numerically by extracting
the coefficient $F_1(k_F,\lambda)$ and verifying the
estimate (\ref{epsilon-estimate}) for the deviation from the
fit. As a result of the fit, we find that the 
coefficient $F_1(k_F,\lambda)$ is purely imaginary and an odd function
of $\lambda$. Moreover, we observe that, to a very high precision, 
$F_1(k_F,\lambda)$ can be described by a simple formula,
\begin{equation}
	F_1(k_F,\lambda)= -\, \frac{i}{4} \left(\frac{\lambda}{\pi}\right)^3  \cot k_F
 \label{F1-formula}
\end{equation}
(see \ref{app:num} for details of the numerical calculations).
We therefore conjecture that this formula is analytically exact
for our expansion (\ref{chi-sum-2})--(\ref{epsilon-estimate}).

To support this conjecture, we have calculated the second and the third
cumulants to the order $L^{-1}$ (see \ref{app:cumulants}).
While $\langle\langle Q^2 \rangle\rangle$ does not produce 
any contribution at the order $L^{-1}$, the third cumulant, to
the leading order, is given by
\begin{equation}
\langle\langle Q^3 \rangle\rangle \sim \frac{3}{2\pi^3 L} \cot k_F\, ,
\label{3-cumulant}
\end{equation}
which is consistent with Eq.~(\ref{F1-formula}). Furthermore, if
there are corrections to Eq.~(\ref{F1-formula}), then they are
of the order higher than $\lambda^3$. Since the relative deviation
of the numerically extracted $F_1(k_F,\lambda)$ from the formula
(\ref{F1-formula}) does not exceed $10^{-4}$ to $10^{-5}$ in the
range of $\lambda = 0.3 \ldots 1.3$ (error bars are higher for $\lambda$
close to zero and to $\pi$), we conclude that the numerical
coefficients at such corrections, if nonzero, would 
be below $10^{-4}$ (at least, for the values of $k_F$ tested).
It seems therefore likely that such corrections vanish
identically and our suggested expression (\ref{F1-formula})
is exact. It would be interesting to verify or disprove this conjecture
by the methods of Ref.~\cite{2010-DeiftItsKrasovsky}, and we leave this for future studies.

\section{Summary and discussion}
\label{sec:summary}

The results of this work are twofold. First, we apply the
recently proven generalized Fisher--Hartwig conjecture to
describe the asymptotic behavior of the FCS problem of
counting free fermions on a line segment. Second, we analyze
numerically the precision of the generalized Fisher--Hartwig
formula and find that, at least in our case, it is more accurate than rigorously
proven. Furthermore, we numerically extract the next-order
correction to the generalized Fisher--Hartwig formula and
conjecture a simple analytical expression for it.

We hope that our results will be useful in several respects.
On the mathematical side, we hope that they will stimulate
further studies of the asymptotic behavior of Toeplitz
determinants. In particular, it seems plausible that our
formulas (\ref{chi-sum-2}) and (\ref{F1-definition}) are
the beginning of a more general \textit{exact} expansion
in powers of $L^{-1}$, see Eqs.\ (\ref{chi-full-sum})
and (\ref{chi-star}) of \ref{app:cumulants-gFH}.
Another interesting question would be
extending the Fisher--Hartwig prescription (\ref{chi-sum})
to {\em block-Toeplitz} determinants \cite{1976-Widom,2007-ItsJinKorepin,2007-BasorEhrhardt}. Such an extension
would be relevant for various FCS problems with noninteracting 
fermions, where the generating function is given by
the Levitov--Lesovik determinant formula  
\cite{1993-LevitovLesovik,1997-IvanovLeeLevitov} related to
block-Toeplitz matrices.

On the physical side, the example we consider in this
work is probably the simplest quantum example of the
``nonanalytic'' FCS phase, as defined in Ref.~\cite{2010-IvanovAbanov},
where the thermodynamic limit ($L\to\infty$) can be 
studied in detail. Remarkably, the ``sum prescription''
(\ref{chi-sum}), which describes the development of
the singularity in the generating function, is the same
as in FCS phase transitions in classical Markov processes
(see, e.g., the discussion of the ``weather model'' in
Ref.~\cite{2010-IvanovAbanov}). It would thus be interesting to explore the
extent of universality of the formula (\ref{chi-sum}),
in particular in the case of interacting quantum models.

\section{Acknowledgments}

We have  benefited from discussions
with A.~Mirlin, D.~Aristov, A.~Its, T.~Giamarchi, L.~Levitov,  and E.~Sukhorukov. 

A.G.A. is grateful to ITP, EPFL for hospitality in summers 2010 and 2011.  
A.G.A. was supported by the NSF under the grant DMR-0906866.

\appendix

\section[\hspace{2.3cm}Bosonization calculation...]{Bosonization calculation of FCS for free fermions}
\label{app:bosonization}
Here we briefly derive (\ref{chiGauss}) (together with (\ref{Q1}) and (\ref{Q2cont})) 
using the bosonization technique (see, e.g., Ref.~\cite{Stone-bosonization}). 
In the simplest version of the bosonization approach, the density of one-dimensional fermions
is given by
\bea
	\rho(x) \approx \rho_{0}-\frac{1}{\pi}\partial_{x}\phi(x),
 \la{rbos-0}
\eea
where $\phi(x)$ is a free bosonic field with the correlation function
\be
	\langle [\phi(x)-\phi(0)]^{2}\rangle
	\approx \ln\frac{|x|}{l_{0}}
 \la{corbos-0}
\ee
with $l_{0}$ being an ultraviolet cutoff (of the order of lattice constant). The expression (\ref{corbos-0}) is good at $|x|\gg l_{0}$.  The number of fermions on a segment of a length $L$ is obtained from (\ref{rbos-0}) as
\be
	Q=\int_{0}^{L}dx\,\rho(x) = \rho_{0}L -\frac{1}{\pi}\Big(\phi(L)-\phi(0)\Big).
 \la{Qbos}
\ee
Upon substituting this bosonized form in (\ref{FCSdef}), we find
\bea
	\chi(\lambda) &=& e^{i\lambda\rho_{0}L}
	\Big\langle e^{-i\frac{\lambda}{\pi}(\phi(L)-\phi(0))}\Big\rangle
	=e^{i\lambda\rho_{0}L}
	 e^{-\frac{\lambda^{2}}{2\pi^{2}}\langle(\phi(L)-\phi(0))^{2}\rangle}
 \nonumber \\
	 &=& e^{i\lambda\rho_{0}L} \left(L/l_{0}\right)^{-\lambda^{2}/(2\pi^{2})}\, ,
 \la{chibos-0}
\eea
which gives us (\ref{chiGauss}). 

This calculation misses the most important property of $\chi(\lambda)$: its periodicity
in $\lambda$. This is due to the approximation (\ref{rbos-0}), which neglects
terms oscillating with the wavevector $2k_{F}$ and its multiples (and thus carries
no information about the particle discreteness). The effect of such terms was
taken into account in Ref.~\cite{1998-Aristov} at a phenomenological level, which
resulted in a qualitatively (but not quantitatively) correct approximation. The
ansatz of Ref.~\cite{1998-Aristov} can be also reproduced in a different way: suppose
we calculate the probabilities of different particle numbers as the Fourier transform of the
bosonization result (\ref{chibos-0}) and then restrict the particle number to
be integer. This prescription produces the particle-number probabilities
\begin{equation}
	P_{q} \propto e^{-\frac{(q-\langle Q\rangle)^{2}}{2\langle\langle Q^{2}\rangle\rangle}}\, ,
\end{equation}
which result in the generating function
\be
	\chi(\lambda) = \sum_{q=-\infty}^{+\infty}P_{q}e^{i\lambda q}
	\propto \sum_{j=-\infty}^{+\infty}
	e^{i(\lambda-2\pi j)\langle Q\rangle 
	-\frac{(\lambda-2\pi j)^{2}}{2}\langle\langle Q^{2}\rangle\rangle}\, ,
 \label{per-chi}
\ee
where we used Poisson's summation formula (the overall coefficient being determined 
from the normalization condition $\chi(\lambda=0)=1$).
Remarkably, this ansatz coincides with the Aristov's conjecture in Ref.~\cite{1998-Aristov}.
It is only qualitatively correct (the errors being of relative order $O(L^{0})$), as
can be seen from the exact results  based on the Fisher--Hartwig formula.

\section[\hspace{2.3cm}Fisher--Hartwig formula...]{Fisher--Hartwig formula for FCS of free fermions}
\label{app:FH}

In this Appendix, we review the Toeplitz-determinant approach to the problem
and apply the generalized Fisher--Hartwig formula.

\subsection[\hspace{2.3cm}Determinant representation...]{Toeplitz-determinant 
representation for $\chi(\lambda)$}
\la{sec:Td}

We start by reproducing the Toeplitz-determinant representation of $\chi(\lambda)$ following \cite{LSM-1961}. 
We consider \textit{lattice} fermions hopping on an infinite one-dimensional lattice with sites labeled by integer numbers. The single-particle states are parameterized
by the wave vector $k$ from the Brillouin zone, $k\in [-\pi,\pi]$. We assume that
the ground state consists of the filled states with wavevectors $-k_{F}<k<k_{F}$, 
where $k_{F}$ is the Fermi wavevector. We do not specify Hamiltonian here, as we are interested in the static problem, where the correlator (\ref{FCSdef}) is determined only by the ground state of the system. The one-particle Green's function in the coordinate space is given by
\be
	g_{ij}=g_{i-j} \equiv \langle c_{i}^{\dagger}c_{j}\rangle = 
        \int_{-\pi}^{\pi}\frac{dk}{2\pi}\, g(k) e^{-ik(i-j)},
 \la{Gij1}
\ee
where the Green's function in momentum space is the step function equal to $1$ for $|k|<k_{F}$
and $0$ otherwise:
\be
	g(k) = \Theta(k_{F}-|k|)
 \la{Gk}
\ee
Explicitly combining (\ref{Gij1}) and (\ref{Gk}), we obtain
\be
	g_{i-j} =  \frac{\sin(k_{F}(i-j))}{\pi(i-j)}
 \la{Gij}
\ee
with $g_{0}=k_{F}/\pi$.
The operator of the number of fermions on the segment $1\leq j\leq L$ is given by
\be
	Q = \sum_{i=1}^{L}c_{i}^{\dagger}c_{i}.
 \la{Qlat}
\ee
Then we can rewrite (\ref{FCSdef}) as
\bea
	\chi(\lambda) = \left\langle e^{i\lambda  \sum_{i=1}^{L}c_{i}^{\dagger}c_{i}}\right\rangle
	=\left\langle \prod_{i=1}^{L} (1+(e^{i\lambda}-1)c_{i}^{\dagger}c_{i}) \right\rangle.
 \la{chi1}
\eea
Here we used the projector property $(c_{i}^{\dagger}c_{i})^{2}=c_{i}^{\dagger}c_{i}$. Applying the Wick theorem to (\ref{chi1}) we obtain
\be
	\chi(\lambda) =\det\bm{T}_{L}= \det(\bm{1}+(e^{i\lambda}-1)\bm{g})_{L\times L},
 \la{Tdet}
\ee
where $\bm{1}$ is the unit $L\times L$ matrix and $\bm{g}$ is the $L\times L$ matrix with 
the matrix elements $g_{ij}=g_{i-j}$ given by (\ref{Gij}).

The determinant (\ref{Tdet}) is that of the Toeplitz matrix $\bm{T}_{L}$ (its matrix 
elements $(T_{L})_{ij}$ depend only on the difference of the row and column indices $i-j$). 
It is said that this Toeplitz matrix $\bm{T}_{L}$ has the \textit{symbol} 
$f(e^{i\theta})$ with 
\be
	f(e^{i\theta}) = 1+(e^{i\lambda}-1)\Theta(k_F-|\theta|)\, ,
 \la{sigma}
\ee
so that the matrix elements are given by
\be
	(T_{L})_{ij} = \int \frac{d\theta}{2\pi}\, e^{iq(i-j)} f(e^{i\theta}).
\ee
The formulas (\ref{Tdet}) and (\ref{Gij}) express the FCS $\chi(\lambda)$ 
as the determinant of a given $L\times L$ Toeplitz matrix. As we are interested in the limit 
of large length of a segment, we need to find the asymptotics of the Toeplitz determinant 
(\ref{Tdet}) as $L\to \infty$.

\subsection[\hspace{2.3cm}FH conjecture]{The Fisher--Hartwig conjecture}
\la{sec:fh}

Let us calculate the asymptotics of the Toeplitz determinant (\ref{Tdet}) as $L\to\infty$ 
using the Fisher--Hartwig conjecture \cite{FisherHartwig-1968,2010-DeiftItsKrasovsky}. This Appendix is not self-contained. We explicitly 
refer to the notations and the results of Ref.~\cite{2010-DeiftItsKrasovsky} (except
for the matrix size denoted $L$ in our paper and $n$ in Ref.~\cite{2010-DeiftItsKrasovsky}). \footnote{For a recent similar application of the Fisher-Hartwig conjecture see \cite{2005-FranchiniAbanov,2011-GutmanGefenMirlin}.}

The symbol (\ref{sigma}) of the Toeplitz matrix has two singularities. In the notation
of Ref.~\cite{2010-DeiftItsKrasovsky}, one of the singularities must be at $\theta=0$,
so if one follows the notation of that paper, one needs to consider an equivalent
problem with the symbol
\be
	f(e^{i\theta}) = \left\{\begin{array}{ll}
					e^{i\lambda} & \;\;\mbox{for  }\;\; 0< \theta <2k_{F} \\
					1                  & \;\;\mbox{for  }\;\; 2k_{F}< \theta< 2\pi
					\end{array}
				\right.
 \label{fsymbol}
\ee

The two point singularities (Fisher-Hartwig singularities) on the unit circle are
located at $z_{0}=1$ and $z_{1}=e^{2ik_{F}}$. These are pure phase discontinuities 
characterized by $\alpha_{0}=\alpha_{1}=0$ and $\beta_{1}=-\beta_{0}=\lambda/2\pi$ 
(for notations see Ref.~\cite{2010-DeiftItsKrasovsky}). The regular part of 
the symbol $f(z)$ is given by $V_{k}=0$ for $k\neq 0$ and by $V_{0}=i2k_{F}\frac{\lambda}{2\pi}$. Then the theorem 1.1 of \cite{2010-DeiftItsKrasovsky} (due to Ehrhardt \cite{Ehrhardt-2001}) 
gives
\bea
	\chi(\lambda) &\sim& e^{2ik_{F}\kappa_{0}L}
	(2L\sin k_{F})^{-2\kappa_{0}^{2}}  \left[G(1+\kappa_{0})G(1-\kappa_{0})\right]^{2}.
 \la{fhresult} 
\eea
Here $\kappa_{0}\equiv \frac{\lambda}{2\pi}$ and the result (\ref{fhresult}) is valid for $-\pi<\lambda<\pi$ in the asymptotic sense as $L\to \infty$. The relative
accuracy of (\ref{fhresult}) is $o(1)$. 

The Barnes $G$ function used in (\ref{fhresult}) is defined as \cite{DLMF}
\be
	G(1+z) \equiv (2\pi)^{z/2}e^{-[z+(\gamma_{E}+1)z^{2}]/2}
	\prod_{k=1}^{\infty}\left(1+\frac{z}{k}\right)^{k}e^{-z+\frac{z^{2}}{2k}},
 \la{GBarnes}
\ee
where $\gamma_{E}\approx 0.57721\ldots$ is the Euler-Mascheroni constant. We have from (\ref{GBarnes})
\be
	G(1+z)G(1-z) =e^{-(\gamma_{E}+1)z^{2}}
	\prod_{k=1}^{\infty}\left(1-\frac{z^{2}}{k^{2}}\right)^{k}e^{\frac{z^{2}}{k}}. 
 \la{GBarnesId}
\ee

To extend the result (\ref{fhresult}) to $\lambda$ outside the interval $(-\pi,\pi)$, one has to introduce the \textit{realizations} of the symbol \cite{2010-DeiftItsKrasovsky}. In our particular case it amounts to using (\ref{fhresult}) with $\kappa_{0}$ replaced by  
\be
	\kappa_{j}= \frac{\lambda}{2\pi}-j
 \la{kappaj}
\ee
for $-\pi+j<\lambda<\pi+j$, $j=0,\pm1, \pm 2,\ldots$. In other words, for a given $\lambda$,
 one should replace in (\ref{fhresult}) $\kappa_{0}\to \kappa_{j}$ with the value $j$ 
minimizing $\kappa_{j}^{2}$. It is important that for $\lambda$ being an odd multiple 
of $\pi$ there are two such values of $j$ minimizing $\kappa_{j}^{2}$. In these cases 
one should replace (\ref{fhresult}) by the sum over corresponding realizations. 
In particular, according to the theorem 1.13 of \cite{2010-DeiftItsKrasovsky} we have 
for $\lambda=\pi$
\bea
	\chi(\lambda) &\sim& \sum_{j=0,1} e^{2ik_{F}\kappa_{j}L}
	(2L\sin k_{F})^{-2\kappa_{j}^{2}}  \left[G(1+\kappa_{j})G(1-\kappa_{j})\right]^{2}.
 \la{gfhresult} 
\eea
Here $\kappa_{0,1}=\pm 1/2$. We argue in the main text that in fact, the accuracy
of this formula is $O(1/L)$. Moreover, the prescription of adding the most relevant
realizations produces correct subleading terms even away from $\lambda=\pi$. 
\footnote{A similar improvement of Fisher-Hartwig formula was shown to work numerically for some Toeplitz determinants in Ref.~\cite{2005-FranchiniAbanov}.}

\subsection[\hspace{2.3cm}Limits $\lambda\to 0$ and $\lambda=\pi$]{Limits 
$\lambda\to 0$ and $\lambda=\pi$}
\la{sec:limits}

For completeness, we present here the two limiting cases: $\lambda\to 0$ and $\lambda=\pi$.

For small $\lambda$, the leading realization in (\ref{gfhresult}) is given by $j=0$ 
(with the term $j=1$ suppressed compared to $j=0$ by about $1/L^{2}$). We have
\bea
	\chi(\lambda\ll \pi) 
 	&\approx& e^{i\lambda \rho_{0}L
	-\frac{\lambda^{2}}{2\pi^{2}}\ln\left[2L e^{\gamma_{E}+1}\sin k_{f}\right]
	-\left(\frac{\lambda}{2\pi}\right)^{4}\zeta(3)}\, ,
 \la{gfhsmalllambda}
\eea
where we have expanded in small $z$,
\be
	G(1+z)G(1-z)\approx e^{-(\gamma_{E}+1)z^{2}-\frac{1}{2}\zeta(3) z^{4}+\ldots}\, .
\ee
We extract from (\ref{gfhsmalllambda})
\bea
	\langle Q\rangle &=&\frac{k_{f}}{\pi}L= \rho_{0}L,
 \\
 	\langle\langle Q^{2}\rangle\rangle & = & 
	\frac{1}{\pi^{2}}\ln\left[2L e^{\gamma_{E}+1}\sin k_{f}\right] +o(1),
 \\
 	\langle\langle Q^{4}\rangle\rangle & = & -\frac{3}{2\pi^{4}}\zeta(3)+o(1).
 \\
 	\langle\langle Q^{2n+1}\rangle\rangle & = & o(1), \;\;n=1,2,\ldots.
\eea

Let us now consider the case $\lambda=\pi$. Then the dominating parametrizations 
are $j=0,1$ (or, equivalently, $\kappa=\pm 1/2$). 
We sum over these parametrizations in (\ref{gfhresult}) and obtain
\be
	\chi(\lambda = \pi)
	\sim 2\left[G\left(1/2\right)G\left(3/2\right)\right]^{2}
	\frac{\cos(k_{F}L)}{\sqrt{2L\sin k_{F}}}.
 \la{gfhpilambda} 
\ee

\section[\hspace{2.3cm}Numerics]{Details of numerical calculations}
\label{app:num} 

Numerical calculations of the Toeplitz determinant (\ref{Tdet}),
up to $L=5000$, were performed using the superfast algorithm of Ref.~\cite{program} implemented in C++ and, independently, 
with the help of Wolfram Mathematica \cite{Wolfram} (the two methods agree within
the error bars, which range from $10^{-9}$ to $10^{-12}$, depending
on the method). For our computations, we took three values of $k_F$
($k_F/\pi= 1/30$, $3/17$, and $8/17$) and various values of $\lambda$
(in multiples of $0.1\pi$). The obtained determinants were then
fitted according to Eq.~(\ref{F1-definition}), and the coefficients
$F_1(k_F,\lambda)$ were extracted for each pair of the parameters
$k_F$ and $\lambda$ used. This fitting procedure involved splitting
the whole range of values of $L$ into intervals of an adjustable
length and fitting within each interval (using $F_1(k_F,\lambda)$
and $F_1(k_F,\lambda-2\pi)$ as the two fitting parameters). Then
the fit parameters were extrapolated to
$L\to\infty$ by using quadratic or cubic polynomials in $L^{-1}$.
This fitting procedure allowed us to obtain a very good precision
for $F_1(k_F,\lambda)$.


\begin{table}
\begin{center}
\tabcolsep=2mm
\begin{tabular}{c|llll}
$\lambda/\pi$ & $k_F=\pi/30$ & $k_F=3\pi/17$ & $k_F=8\pi/17$ \\
\hline
0.1  &  0.002382(9) &  0.0004043(4)     &  0.00002316(3)$^*$ \\
0.2  &  0.01904(1)  &  0.0032315(6)     &  0.0001858(3)      \\
0.3  &  0.06423(1)  &  0.0109030(7)     &  0.0006259(6)      \\
0.4  &  0.15224(1)  &  0.0258409(4)$^*$ &  0.0014840(8)      \\
0.5  &  0.29734(2)  &  0.050472(2)      &  0.0028957(4)$^*$  \\
0.6  &  0.51378(3)  &  0.087217(3)      &  0.0050038(1)$^*$  \\
0.7  &  0.81586(5)  &  0.138495(3)      &  0.007947(1)       \\
0.8  &  1.21783(9)  &  0.206731(4)      &  0.011862(1)       \\
0.9  &  1.7340(2)   &  0.294348(5)      &  0.016889(2)       \\
1.0  &  2.3785(3)   &  0.403763(7)      &  0.023166(3)       \\
1.1  &  3.1658(5)   &  0.53740(1)       &  0.030832(5)       \\
1.2  &  4.1100(9)   &  0.69770(4)       &  0.04003(1)        \\
1.3  &  5.225(2)    &  0.8870(1)        &  0.05088(5)        \\
1.4  &  6.525(4)    &  1.1078(3)        &  0.06357(3)$^*$    \\
1.5  &  8.025(9)    &  1.362(2)         &  0.0782(2)$^*$     \\
1.6  &  9.74(3)     &  1.654(1)$^*$     &  0.095(4)          \\
1.7  &  11.6(3)     &  1.99(2)          &  0.12(2)           \\
1.8  &  13(3)       &  2.3(1)           &  ---               \\
1.9  &  ---         &  ---              &  ---               \\
\end{tabular}
\end{center}
\caption{Numerical values of $i F_1(k_F,\lambda)$  extracted
from the fits. For most of the data, the determinants were calculated
with the precision $10^{-9}$. For several sets of data (marked by
the asterisks), a higher precision ($10^{-12}$) was used, with the
help of Mathematica \cite{Wolfram}. Using higher precision
results in reducing the error bars for the fitting parameters.
The dashes in the last lines of the table correspond to the
values of $\lambda$, for which $F_1(k_F,\lambda)$ could not
be reliably determined from our computations.}
\label{table:F1}
\end{table}

The resulting values of $F_1(k_F,\lambda)$ are presented in 
Table \ref{table:F1}. Within the error bars, $F_1(k_F,\lambda)$
is purely imaginary, which, in combination with the reality
condition $F_1(k_F, -\lambda) = F_1(k_F,\lambda)^*$, implies that
$F_1(k_F,\lambda)$ is odd in $\lambda$.

\begin{figure}
\begin{center}
	\includegraphics[width=0.80\textwidth]{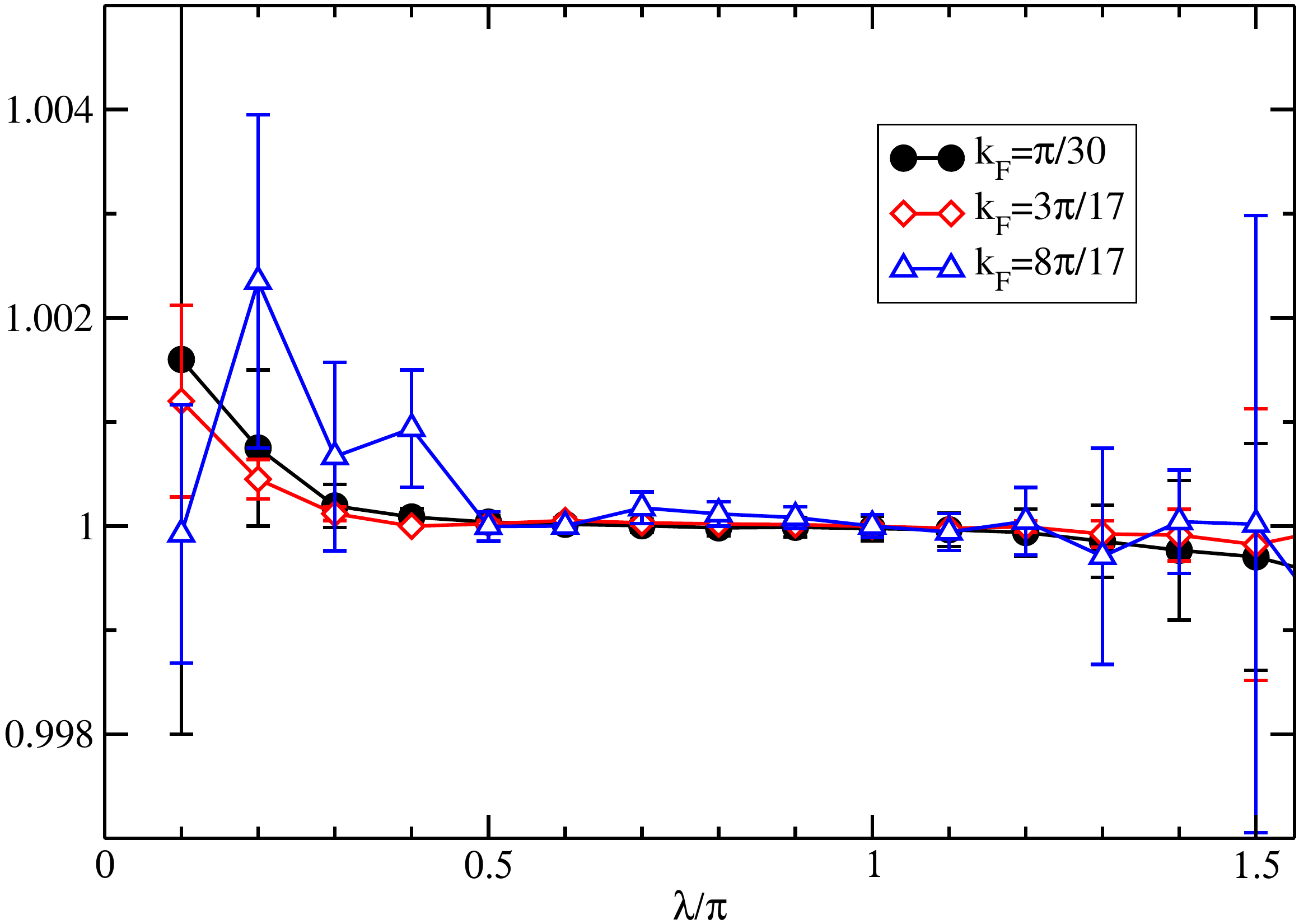}
\end{center}
\caption{The ratio of the numerically found value of $F_1(k_F,\lambda)$
(reported in Table~\ref{table:F1}) to the analytical conjecture 
(\ref{F1-formula}).}
\label{fig:F1}
\end{figure}

We further observe that thus extracted function $F_1(k_F,\lambda)$
can be described by the formula (\ref{F1-formula}) to a very
high precision. The ratio of the numerically extracted 
$F_1(k_F,\lambda)$ to the analytical conjecture (\ref{F1-formula})
is plotted in Fig.~\ref{fig:F1}.

\begin{figure}
\begin{center}
	\includegraphics[width=0.80\textwidth]{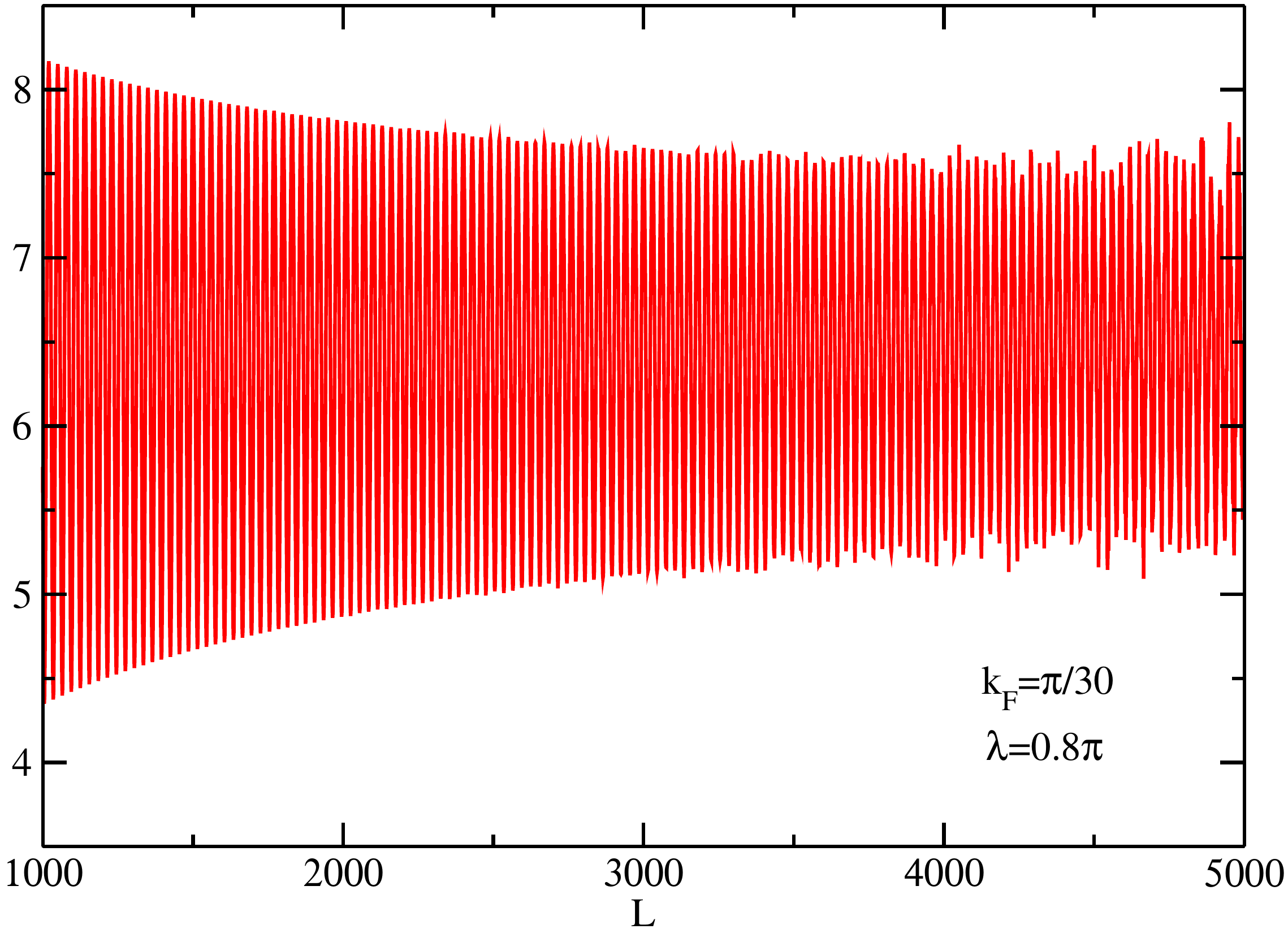}
\end{center}
\caption{A plot of $|\varepsilon|\, L^2 \Big( |\chi_1(\lambda)| + 
|\chi_1(\lambda - 2\pi)| \Big)^{-1}$ for $k_F=\pi/30$ and $\lambda/\pi=0.8$. 
Here $\varepsilon$ is the deviation from our conjectured formula, as defined
by Eqs.~(\ref{chi-sum-2}) and (\ref{F1-definition}), with the
values of $F_1(k_F,\lambda)$ obtained from the numerical fit
(as reported in Table~\ref{table:F1}). The plotted quantity
remains finite in the limit $L\to\infty$, which
supports our estimate (\ref{epsilon-estimate}).}
\label{fig:epsilon}
\end{figure}

Finally, we calculate the error $\varepsilon$ of our expansion at
the order $L^{-1}$, as defined in Eq.~(\ref{chi-sum-2}). 
In Fig.~\ref{fig:epsilon}, we show a typical plot of the ratio
$|\varepsilon|\, L^2 \Big( |\chi_1(\lambda)| + 
|\chi_1(\lambda - 2\pi)| \Big)^{-1}$. The plot suggests that
this quantity remains finite at $L\to\infty$, which supports
our conjecture (\ref{epsilon-estimate}).

In principle, the precision of our numerics should be sufficient
for continuing the expansion in $L^{-1}$ (the next term being of
the order $L^{-2}$). This possibility is also connected to the
question about the precision of the ``sum prescription''
(\ref{chi-sum}): whether this decomposition into a sum 
(involving all shifts of $\lambda$ by multiples of $2\pi$) is exact 
to all (perturbative) orders in $L$ (see \ref{app:cumulants-gFH})
or breaks down at a certain order. 
We leave this interesting mathematical question for future studies.

\section[\hspace{2.3cm}Second and third cumulants...]{Second and third cumulants of the fermionic number on a line segment}
\label{app:cumulants}

In this Appendix, we calculate analytically $\langle\langle Q^2 \rangle\rangle$ and
$\langle\langle Q^3 \rangle\rangle$ up to the orders $L^{-2}$ and $L^{-1}$, respectively. The Wick theorem
expresses those cumulants in terms of the Green function (\ref{Gij1}).
We further denote the set of lattice sites belonging to the segment considered
by $[L]=\{1,2,\dots,L\}$.

For the second cumulant, one gets
\begin{equation}
	\langle\langle Q^2 \rangle\rangle = \sum_{i,j \in [L]} g_{ij} \bar{g}_{ij}
\end{equation}
where
\begin{equation}
	\bar{g}_{ij}=\langle c_i c^\dagger_j \rangle = \delta_{ij} - g_{ji}
\end{equation}
Using the relation
\begin{equation}
	\sum_{j=-\infty}^{\infty} g_{ij} g_{jk} = g_{ik}\, ,
 \label{gf-convolution}
\end{equation}
one can reexpress
\begin{equation}
	\langle\langle Q^2 \rangle\rangle = \sum_{i \in [L] \atop \bar{j} \notin [L]}
	g_{i\bar{j}} g_{\bar{j}i}\, .
\end{equation}
Collecting together all terms with the same distances between the pair
of sites, we arrive at
\begin{equation}
	\langle\langle Q^2 \rangle\rangle = 2 \sum_{x=1}^{\infty} [g_x]^2 s(x)\, ,
\end{equation}
where we define the function $s(x)=\min(x,L)$. Using the explicit expression
(\ref{Gij}), we find 
\bea
	\langle\langle Q^2 \rangle\rangle &=& \frac{1}{\pi^2} \ln L  
	+ \frac{1}{\pi^2} \left[ \gamma_E + 1 + \ln (2\sin k_F)\right] 
\nonumber \\
	&+& \frac{1}{12\pi^2 L^2} - \frac{1}{\pi^2}\frac{\cos(2k_{F}L)}{(2L\sin k_{F})^{2}}
        + O(L^{-3})
\label{Q2-analytic}
\eea
(note that there are no terms of the order $L^{-1}$).

Repeating the same procedure for the third cumulant, we find
\begin{equation}
	\langle\langle Q^3 \rangle\rangle = \sum_{i,j,k \in [L]} 
	(g_{ik} \bar{g}_{ij} \bar{g}_{jk} - g_{ij} g_{jk} \bar{g}_{ik})\, ,
\end{equation}
which, using relation (\ref{gf-convolution}), can be converted to
\begin{equation}
	\langle\langle Q^3 \rangle\rangle = 
	\sum_{i \in [L] \atop \bar{j},\bar{k} \notin [L]}
	g_{i\bar{j}} g_{\bar{j}\bar{k}} g_{\bar{k}i}
	-
	\sum_{i,k \in [L] \atop \bar{j} \notin [L]}
	g_{i\bar{j}} g_{\bar{j}k} g_{ki}\, .
\end{equation}
Again, collecting together all terms with the same relative positions
of points, we arrive at
\begin{equation}
	\langle\langle Q^3 \rangle\rangle
	= 6 \sum_{x,y=1}^{\infty} g_x g_y g_{x+y} 
	\left[s(x)+s(y)-s(x+y)\right]\, .
\end{equation}
Note that there is no contribution to this sum from small $x$ and $y$,
for which $x+y\leq L$. At large $L$, the main contribution to this
sum is determined by the two boundary pieces: ($y=0$, $x>L$) and
($x=0$, $y>L$). The easiest technique to extract these contributions
is summation ``by parts''. If we re-factorize
\begin{equation}
	\langle\langle Q^3 \rangle\rangle
	= 6 \sum_{x,y > 0 \atop x+y>L} \big[ xy\, g_x g_y g_{x+y} \big] 
	\cdot \frac{s(x)+s(y)-s(x+y)}{xy}\, ,
\end{equation}
then we can associate the leading contribution to the jumps
of the function $[s(x)+s(y)-s(x+y)]/(xy)$ at the boundary.
The contribution from the two pieces of the boundary are
equal, and, after a simple calculation, one arrives at
\begin{eqnarray}
	\langle\langle Q^3 \rangle\rangle
	&\approx& \frac{3}{\pi^3} \sum_{x>L \atop y\ge 0}  
	\frac{\sin(2 k_F y)}{x(x+y)} 
 \nonumber \\
	&\approx& \frac{3}{\pi^3}
	\sum_{y=0}^{\infty} \sin(2k_F y)\cdot \int_L^{\infty} \frac{dx}{x^2}
	= \frac{3}{2\pi^3 L} \cot{k_F}\, ,
\label{Q3-analytic}
\end{eqnarray}
to the leading order in $L$
(in the calculation, we neglected the terms oscillating in 
$x$ in $g_x g_y g_{x+y}$, as they produce contributions 
of higher orders).

\section[\hspace{2.3cm}Charge cumulants from the generalized Fisher--Hartwig formula]{Charge cumulants from the generalized Fisher--Hartwig formula}
\label{app:cumulants-gFH}

In this Appendix, we try to push our conjecture about the precision of the
generalized Fisher--Hartwig formula even further than in the main body of
the paper. Namely, we assume that the generalized Fisher--Hartwig ``sum prescription''
is exact to all orders of $L^{-1}$, if all possible \textit{realizations} 
(\ref{kappaj}) are included:
\begin{equation}
	\chi(\lambda) = \sum_{j=-\infty}^{+\infty} \chi_* (\lambda-2\pi j)\, ,
 \label{chi-full-sum}
\end{equation}
where
\begin{equation}
	\chi_*(\lambda) = \exp \left[ i\lambda \frac{k_F}{\pi} L - 
	\frac{\lambda^2}{2\pi^2} \ln\frac{L}{l_0}
	+ \sum_{m=0}^{\infty} F_m(k_F,\lambda) L^{-m} \right]
 \label{chi-star}
\end{equation}
(we assume that the sum in the exponent contains only $L^{-m}$, but
no logarithms). The series (\ref{chi-full-sum}) may be, in general,
divergent at a fixed $L$ and should be understood as an asymptotic expansion
at $L\to \infty$. Such a conjecture may, in principle, be verified
by comparing with the charge cumulants (\ref{cumdef}). In particular, for
the first several cumulants, we find from Eqs.\ (\ref{chi-full-sum}) and
(\ref{chi-star})\footnote{Here $F_m^{(k)}$ 
means $\partial_{\lambda}^{k}F_m\Big|_{\lambda=0}$}:
\bea
	\langle Q\rangle &=& \frac{1}{\pi}k_{F}L \, ,
 \\
 	\langle\langle Q^{2}\rangle\rangle &=&
	\frac{1}{\pi^{2}}\ln(e^{\gamma_{E}+1}2L\sin k_{F})
	-\frac{1}{\pi^{2}}\frac{\cos(2k_{F}L)}{(2L\sin k_{F})^{2}}
\nonumber \\
	& & - F_{1}'' L^{-1} - F_{2}'' L^{-2} + o(L^{-2}) \, ,
 \\
 	\langle\langle Q^{3}\rangle\rangle &=&
	+\frac{6}{\pi^{3}}\frac{\sin(2k_{F}L)}{(2L\sin k_{F})^{2}}
	\ln(e^{\gamma_{E}}2L\sin k_{F})
\nonumber \\
	& & + i F_{1}^{(3)} L^{-1} + i F_{2}^{(3)} L^{-2} + o(L^{-2}) \, ,
 \\
 	\langle\langle Q^{4}\rangle\rangle &=&
	-\frac{3}{2\pi^{4}}\zeta(3)
	+\frac{24}{\pi^{4}}\frac{\cos(2k_{F}L)}{(2L\sin k_{F})^{2}}
	\left[\ln(e^{\gamma_{E}}2L\sin k_{F})\right]^{2}
 \nonumber \\
	& & + F_{1}^{(4)} L^{-1}+F_{2}^{(4)} L^{-2} + o(L^{-2})\, .
\eea	
In other words, the coefficients in the expansions of the cumulants
$\langle\langle Q^n\rangle\rangle$ in $L^{-1}$ may be related to the
expansions of $F_m(k_F,\lambda)$ in $\lambda$ (at $\lambda=0$). These
relations may, in principle, be verified against direct calculations
of the cumulants using the Wick theorem (along the lines of \ref{app:cumulants}).

In the present work, we have only calculated analytically the cumulant
$\langle\langle Q^2\rangle\rangle$ to the order $L^{-2}$ and
the cumulant $\langle\langle Q^3\rangle\rangle$ to the order $L^{-1}$
[see Eqs.\ (\ref{Q2-analytic}) and (\ref{Q3-analytic}), respectively].
This implies $F_{1}''=0$, $F_{2}''=-1/(12\pi^2)$, 
and $F_{1}^{(3)}=-(3i/2\pi^3 L) \cot{k_F}$
[in agreement with our conjecture (\ref{F1-formula})]. Furthermore,
numerical studies of the third and fourth cumulants indicate 
that $F_{2}^{(3)}=F_{1}^{(4)}=0$  and $F_2^{(4)}\ne 0$. In other words,
if our conjecture in Eqs.\ (\ref{chi-full-sum}) and (\ref{chi-star})
is correct, then $F_2(k_F,\lambda)$ has the form
\begin{equation}
F_2(k_F,\lambda) = -\frac{\lambda^2}{24 \pi^2} + O(\lambda^4)
\end{equation}
in its expansion around $\lambda=0$. It seems likely that further
analytical progress is possible in these questions, and we leave
them for future studies.

\section*{References}


\end{document}